\documentclass[conference]{IEEEtran}
\usepackage{amsthm}
\usepackage{times,amssymb,amsmath,amsfonts,float,nicefrac,color,bbm,mathrsfs,float}
\usepackage{booktabs}
\usepackage{hyperref}
\usepackage{algorithm,enumerate,multirow,caption,tikz,graphicx}
\usepackage{algpseudocode}
\usepackage[noadjust]{cite}
\usepackage{subcaption}
\usepackage{comment}

\allowdisplaybreaks

\newcommand{\RN}[1]{%
  \textup{\expandafter{\romannumeral#1}}%
}

\tikzset{
  block/.style    = {draw, thick, rectangle, minimum width = 3em},
  sblock/.style      = {draw, thick, rectangle, minimum height = 3em,
    minimum width = 3em}, 
}

\newcommand\remove[1]{}

\allowdisplaybreaks

\newtheorem{cnstr}{Construction}

\newcommand{\cC}{\mathcal{C}}

\DeclareMathOperator{\out}{out}
\DeclareMathOperator{\lcm}{lcm}

\begin{document}
\title{Improving the List Decoding Version of the Cyclically Equivariant Neural Decoder}

\author{Xiangyu Chen \and Min Ye}

\maketitle
{\renewcommand{\thefootnote}{}\footnotetext{

\vspace{-.2in}
 
\noindent\rule{1.5in}{.4pt}

X. Chen is with Tsinghua-Berkeley Shenzhen Institute, Tsinghua Shenzhen International Graduate School, Shenzhen 518055, China.

M. Ye is with Tsinghua-Berkeley Shenzhen Institute, Tsinghua Shenzhen International Graduate School, Shenzhen 518055, China. Email: yeemmi@gmail.com
}

\renewcommand{\thefootnote}{\arabic{footnote}}}
\setcounter{footnote}{0}

\begin{abstract}
The cyclically equivariant neural decoder was recently proposed in [Chen-Ye, {\em International Conference on Machine Learning}, 2021] to decode cyclic codes. In the same paper, a list decoding procedure was also introduced for two widely used classes of cyclic codes---BCH codes and punctured Reed-Muller (RM) codes. While the list decoding procedure significantly improves the Frame Error Rate (FER) of the cyclically equivariant neural decoder, the Bit Error Rate (BER) of the list decoding procedure is even worse than the unique decoding algorithm when the list size is small. In this paper, we propose an improved version of the list decoding algorithm for BCH codes and punctured RM codes. Our new proposal significantly reduces the BER while maintaining the same (in some cases even smaller) FER. More specifically, our new decoder provides up to $2$dB gain over the previous list decoder when measured by BER, and the running time of our new decoder is $15\%$ smaller. Code
available at \href{https://github.com/improvedlistdecoder/code}{github.com/improvedlistdecoder/code}
\end{abstract}

\section{Introduction}\label{sect:intro}
Machine learning methods have recently been applied to the area of decoding error-correcting codes \cite{Nachmani16,Gruber17,Cammerer17,Nachmani18,Kim18,Kim18a,Vasic18,Teng19,Jiang19,Nachmani19,Carpi19,Habib20,Buchberger20}. These methods have demonstrated improvements over the classical decoding algorithms for codes with short to moderate block length. In particular, one line of research pioneered by \cite{Nachmani16,Nachmani18} introduced neural decoders as a generalization of the classic Belief Propagation (BP) decoding algorithm, where the Trellis graph in the BP algorithm is viewed as a fully connected neural network \cite{Nachmani16}, and the weights in the Trellis graph are optimized by training the neural network. The fully connected neural networks were further replaced by recurrent neural networks (RNNs) in \cite{Nachmani18}, and the tools of graph neural networks were also introduced to improve the neural decoders \cite{Nachmani19}.

Very recently, the cyclically equivariant neural decoder was proposed to decode cyclic codes \cite{Chen21}. Inspired by the fact that the Maximum Likelihood (ML) decoder of any cyclic code is equivariant to cyclic shifts, \cite{Chen21} imposed a shift invariant structure on the weights of the neural decoder so that it shares the equivariant property of the ML decoder. More precisely, any cyclic shift of inputs results in the same cyclic shift of the decoding outputs for the cyclically equivariant neural decoder. Simulations with BCH codes and punctured Reed-Muller (RM) codes demonstrated that the cyclically equivariant neural decoder consistently outperforms the conventional neural decoders when decoding cyclic codes \cite{Chen21}.

In addition to the cyclically equivariant neural decoder, \cite{Chen21} further proposed a list decoding procedure for BCH codes and punctured RM codes which significantly improves the Frame Error Rate (FER) of the cyclically equivariant neural decoder. For certain high-rate BCH codes and punctured RM codes, the list decoder with a large enough list size achieves almost the same FER as the ML decoder. However, the Bit Error Rate (BER) of the list decoding procedure in \cite{Chen21} is even worse than the unique decoding algorithm when the list size is small. This can be explained as follows: FER is the fraction of incorrectly decoded codewords, and BER is the fraction of incorrectly decoded bits (or codeword coordinates). We say that a codeword is incorrectly decoded whenever the decoding result is different from the true codeword, no matter it differs in all coordinates or it differs only in a single bit. Therefore, the fraction of incorrectly decoded bits in the incorrectly decoded codewords does not affect FER at all but it is an important factor of BER. In fact, BER is simply the product of FER and the average fraction of incorrectly decoded bits in the incorrectly decoded codewords. The major downside of the list decoder in \cite{Chen21} is that it does not optimize this fraction at all. As a consequence, whenever the list decoder makes a mistake, the decoding result differs from the true codeord in more or less half of all the codeword coordinates. On the other hand, although the unique decoding version of the cyclically equivariant neural decoder has a larger FER, it tries to minimize the Hamming distance between the decoding results and the true codewords even when it can not completely recover the true codeword. That's why it has an even better BER compared to the list decoder with a small list size.

In this paper, we propose a new neural decoder for BCH codes and punctured RM codes which improves upon the list decoding version of the cyclically equivariant neural decoder. More precisely, it achieves a significantly smaller BER compared to the list decoder in \cite{Chen21} while maintaining the same (in some cases even smaller) FER. Moreover, the running time of our new decoder is also reduced by $15\%$ compared to the list decoder in \cite{Chen21}.

Both our new decoder and the previous list decoder make use of the affine invariant property of extended BCH codes and RM codes. In the previous list decoder, we associate a parity check matrix to each affine permutation on the codeword coordinates, and we use the cyclically equivariant neural decoder to perform neural Belief Propagation on the Tanner graph of each parity check matrix. After obtaining the list of decoding results from all affine permutations, the final decoding result is obtained from the ML decoding among this list. In our new decoder, we build a large parity check matrix containing all the rows of the parity check matrices used in the previous list decoder, and our new decoder performs neural BP on the Tanner graph of this large parity check matrix. Similarly to the cyclically equivariant neural decoder in \cite{Chen21}, the weights in our new decoder also satisfy certain invariant structure, which brings much better performance than the vanilla neural BP decoders.

\section{Background on RM codes and BCH codes}
\label{sect:back}

In this section, we collect some basic properties about (punctured) RM codes and (extended) BCH codes that are needed to develop our new decoder. Readers may consult \cite{Macwilliams77,Abbe21} for more information about these two code families. 

In order to define BCH codes and punctured RM codes, we first introduce some notation. Let $m$ be an integer, and let $\alpha$ be a primitive element of the finite field $\mathbb{F}_{2^m}$. For $1\le j\le 2^m-2$, let $M^{(j)}(x)$ be the minimal polynomial of $\alpha^j$ over the binary field. Both BCH codes and punctured RM codes are cyclic codes, and they can be defined by generator polynomials and parity check polynomials. More specifically, for BCH code with designed distance $2\delta+1$ and code length $n=2^m-1$, the generator polynomial is $g(x)=\lcm\{M^{(1)}(x),M^{(3)}(x),\dots,M^{(2\delta-1)}(x)\}$, where $\lcm$ stands for least common multiple; see Chapter 7.6 of \cite{Macwilliams77}. For $r$th order punctured RM code with code length $n=2^m-1$, the generator polynomial is 
$
g(x)=\lcm\{M^{(j)}(x): 1\le j\le 2^m-2, ~ 1\le w_2(j)\le m-r-1 \},
$
where $w_2(j)$ is the number of $1$'s in the binary expansion of $j$; see Chapter 13.5 of \cite{Macwilliams77}.

For a cyclic code with code length $n$, the generator polynomial $g(x)$ always divides $x^n-1$, and the parity check polynomial is simply $h(x)=(x^n-1)/g(x)$. For an $(n,k)$ cyclic code, the degree of $h$ is $k$, and so $h(x)$ can be written as $h(x)=h_k x^k + \dots + h_2 x^2+ h_1 x + h_0$, where the coefficients $h_k,\dots,h_2,h_1,h_0$ are either $0$ or $1$. The following $(n-k)\times n$ matrix
\begin{equation} \label{eq:generalpc}
\begin{array}{ccccccccc}
h_k & \dots & h_2 & h_1 & h_0 & 0 & 0 &\dots & 0 \\
0 & h_k & \dots & h_2 & h_1 & h_0 & 0 & \dots & 0 \\
\vdots & \vdots & \vdots & \vdots & \vdots & \vdots & \vdots & \vdots & \vdots \\
0 & 0 & \dots & 0 & h_k & \dots & h_2 & h_1 & h_0
\end{array} 
\end{equation}
is a parity check matrix of the cyclic code, and this particular parity check matrix is used in the neural BP decoders in \cite{Nachmani16,Nachmani18} for BCH codes.

RM codes and extended BCH codes are obtained from adding an overall parity bit to punctured RM codes and BCH codes, respectively. More precisely, if $\cC$ is a punctured RM code, then $\{(C_0,C_1,\dots,C_n):(C_1,\dots,C_n)\in\cC, C_0=C_1+\dots+C_n\}$ is a RM code. Similarly, if $\cC$ is a BCH code, then $\{(C_0,C_1,\dots,C_n):(C_1,\dots,C_n)\in\cC, C_0=C_1+\dots+C_n\}$ is an extended BCH code.
It is well known that both RM codes and extended BCH codes are invariant to the affine group \cite{Kasami67}. In order to explain the affine invariant property, we use $(C_0,C_1,\dots,C_n)$ to denote a codeword from an extended BCH code or a RM code with length $n+1=2^m$. Next we define a one-to-one mapping $f$ between the index set $\{0,1,\dots,n\}$ and the finite field $\mathbb{F}_{2^m}=\{0,1,\alpha,\alpha^2,\dots,\alpha^{n-1}\}$ as follows: $f(0)=0$ and $f(i)=\alpha^{i-1}$ for $i\in[n]$.
For $a,b\in\mathbb{F}_{2^m}, a\neq 0$, the affine mapping $X\mapsto aX+b$ defines a permutation on the finite field $\mathbb{F}_{2^m}$, and through the function $f$ it also induces a permutation on the index set $\{0,1,\dots,n\}$. More precisely, for $1\le i\le n$ and $0\le j\le n$, we use $\sigma_{i,j}$ to denote the permutation on $\{0,1,\dots,n\}$ induced by the mapping $X\mapsto f(i)X+f(j)$:
$$
\sigma_{i,j}(v)= f^{-1} \big( f(i) f(v) + f(j) \big) \text{~~for~} v\in\{0,1,\dots,n\}.
$$
The permutations $\{\sigma_{i,j}: 1\le i\le n, 0\le j\le n\}$ form the affine group to which the RM codes and the extended BCH codes are invariant.

The special case $\sigma_{i,0}$ is the permutation that fixes $C_0$ and performs $(i-1)$ cyclic right shifts on $(C_1,C_2,\dots,C_n)$. The extended code is invariant to such a permutation because $(C_1,C_2,\dots,C_n)$ belongs to a cyclic code.

\begin{figure*}
\centering
\includegraphics[width=0.75\textwidth]{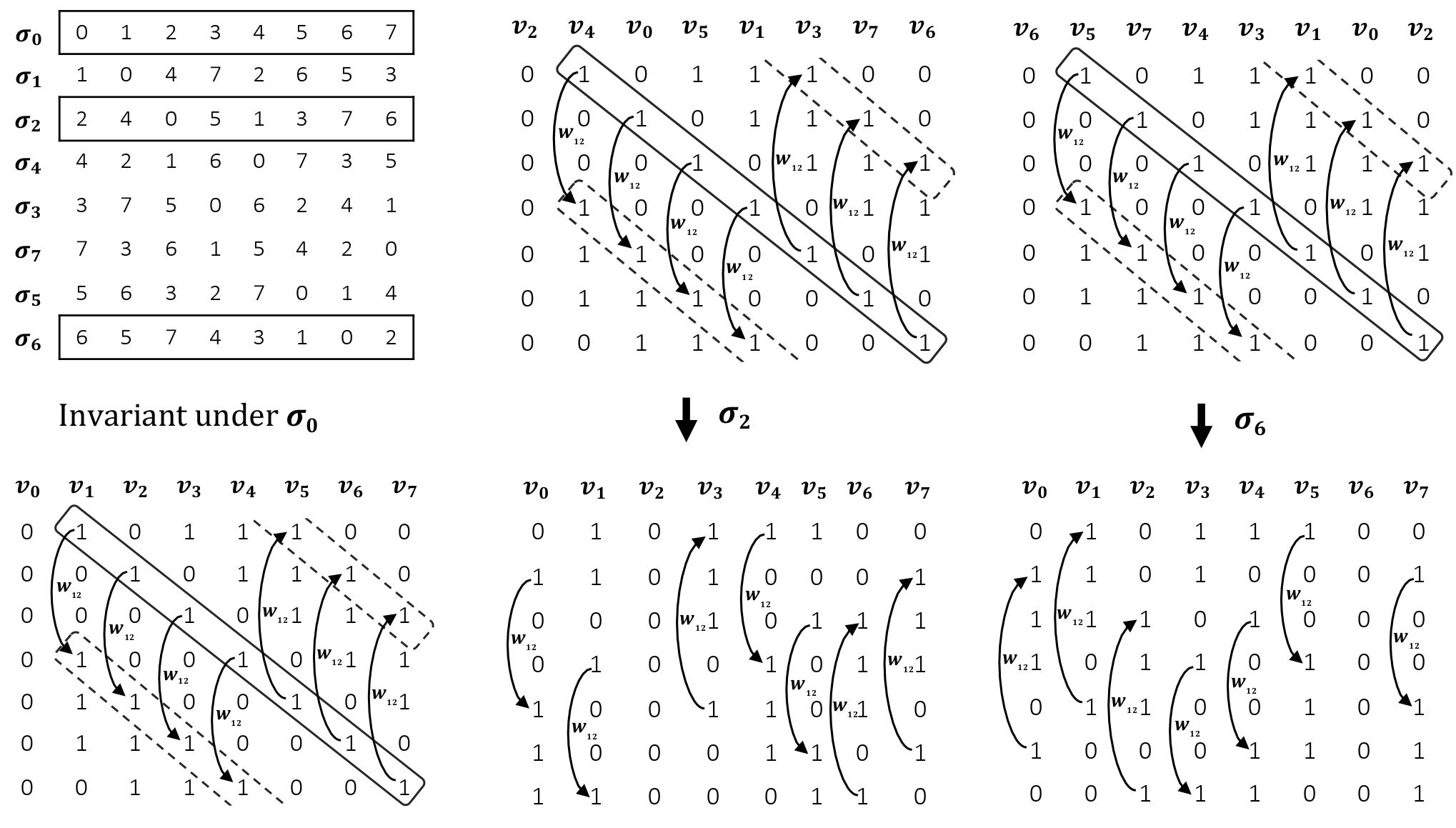}
\caption{$\sigma_0,\dots,\sigma_7$ are the permutations under which the extended $(8,4)$ Hamming code is invariant. In \cite{Chen21} we use the parity check matrix in the bottom-left corner. By applying column permutation $\sigma_j$ to this matrix, we obtain other matrices whose rows are also parity checks of the code, e.g., the two matrices in the bottom-middle and bottom-right corner. In this paper, we build a large parity check matrix $H$ which contains all the row vectors of such matrices, and the weights of our neural decoder are invariant to both the cyclic shifts and the permutations $\sigma_0,\dots,\sigma_7$.}
\label{fig:illus}
\end{figure*}

For both the list decoder in \cite{Chen21} and the new decoder in this paper, we focus on another special case $i=1$, and we write $\sigma_j=\sigma_{1,j}$ to simplify the notation. By definition, $\sigma_j$ is the permutation on $\{0,1,\dots,n\}$ induced by the mapping $X\mapsto X+f(j)$, so $\sigma_j(v)=f^{-1}(f(v)+f(j))$ for $0\le v\le n$. Clearly, $\sigma_0$ is the identity permutation. We will use the set of permutations $\{\sigma_0,\sigma_1,\dots,\sigma_n\}$ in our new decoder. In Fig.~\ref{fig:illus}, we give a concrete example for $n=7$, where each row in the top-left matrix represents a permutation $\sigma_j$.

\section{Our new decoder}
Our new decoder follows the general structure of the neural BP decoders proposed in \cite{Nachmani16,Nachmani18} with some additional structures imposed on the weights. In order to build a BP decoder, we first need to identify a parity check matrix of the code. Typically, a parity check matrix of an $(n,k)$ code has size $(n-k)\times n$. In this case, all the row vectors in the matrix are linearly independent. In our application, however, we allow the number of rows in the parity check matrix to be larger than $n-k$, so the parity check matrix may have some ``redundant" row vectors which are linear combinations of other rows. Suppose that the parity check matrix $H$ has $m$ rows and $n$ columns. The Tanner graph corresponding to $H$ is a bipartite graph constructed as follows: It has $n$ variable nodes labelled as $v_0,v_1,\dots,v_{n-1}$ on the left side and $m$ check nodes labelled as $c_1,c_2,\dots,c_m$ on the right side. An edge is connected between $v_j$ and $c_i$ in the Tanner graph if and only if $H_{ij} = 1$. The inputs of the decoder are the log likelihood ratios (LLRs) of $n$ codeword coordinates: 
$$
L_j = \log \frac{\mathbb{P} (y_j|C_j=0)}{\mathbb{P} (y_j|C_j=1)} 
\quad \text{for~} j\in\{0,1,\dots,n-1\} ,
$$
where $(C_0,\dots,C_{n-1})$ is a randomly chosen codeword, and $(y_0,\dots,y_{n-1})$ is the channel output after transmitting $(C_0,\dots,C_{n-1})$ through $n$ independent copies of some noisy channel. The decoder aims to recover the codeword from the channel output, or equivalently, from the LLRs. In classic BP algorithms, messages propagate back and forth through the edges of the Tanner graph for several iterations. In each iteration, the message on every edge is updated using the messages on its neighboring edges from the previous iteration together with the LLRs. More precisely, in every odd iteration, the message on an edge $(c_i,v_j)$ is updated using the messages on all the other edges that are connected to $v_j$ together with the LLR $L_j$. In every even iteration, the message on an edge $(c_i,v_j)$ is updated using the messages on all the other edges that are connected to $c_i$. The final decoding result of the $j$th coordinate is obtained by summing up $L_j$ and the messages on all the edges connected to $v_j$. In \cite{Nachmani16,Nachmani18}, a set of learnable weights are added into the calculations of odd iterations and the final outputs. In \cite{Chen21}, a cyclically invariant structure is imposed on the learnable weights to obtain better performance when decoding cyclic codes.

We will only describe our decoder for RM codes and extended BCH codes because the decoder for punctured RM codes and BCH codes only requires a trivial modification: If we want to decode the punctured codes, we only need to append a zero entry to the LLR vector. This zero entry means that we know nothing about the overall parity bit. After that the decoder for extended codes can be directly applied to obtain the decoding results. 

Let $\cC$ be a RM code or an extended BCH code with code length $n=2^m$, and let $(C_0,\dots,C_{n-1})$ be a codeword of $\cC$. Without loss of generality, we assume that the last $(n-1)$ coordinates form a cyclic code $\widetilde{\cC}$, which is either a punctured RM code or a BCH code. As mentioned in Section~\ref{sect:back}, a parity check matrix of $\widetilde{\cC}$ has the form \eqref{eq:generalpc}, where every row of this matrix is a cyclic shift of its first row. Due to the cyclically invariant property of $\widetilde{\cC}$, one can show that every cyclic shift of the first row is a parity check of $\widetilde{\cC}$. In total, there are $(n-1)$ cyclic shifts, and we build an $(n-1)\times(n-1)$ parity check matrix of $\widetilde{\cC}$ consisting of these $(n-1)$ row vectors. Finally, by appending an all-zero column vector in front of this $(n-1)\times(n-1)$ matrix, we obtain an $(n-1)\times n$ parity check matrix of the original code $\cC$. As a concrete example, the parity check matrix of the form \eqref{eq:generalpc} for $(7,4)$ Hamming code is
$$
\begin{array}{ccccccc}
1 & 0 & 1 & 1 & 1 & 0 & 0 \\
0 & 1 & 0 & 1 & 1 & 1 & 0 \\
0 & 0 & 1 & 0 & 1 & 1 & 1
\end{array} 
$$
while the matrix in the bottom-left corner of Fig.~\ref{fig:illus} is a parity check matrix of the $(8,4)$ extended Hamming code. Given the code $\cC$, we denote such an $(n-1)\times n$ parity check matrix as $H_0$. Since $\cC$ is invariant under the permutations $\sigma_0,\sigma_1,\dots,\sigma_{n-1}$ defined in Section~\ref{sect:back}, the matrix obtained by performing column permutation $\sigma_j$ on the matrix $H_0$ is also a parity check matrix of $\cC$ for all $0\le j\le n-1$. For example, the matrix in the bottom-middle of Fig.~\ref{fig:illus} is obtained by column permutation $\sigma_2$, and the matrix in the bottom-right corner is obtained by column permutation $\sigma_6$. We use $H_j$ to denote the matrix obtained by column permutation $\sigma_j$.

Our decoding algorithm has a parameter $P\in[n]$, which is the number of permutations we use in our algorithm. By increasing the value of $P$, the algorithm achieves smaller decoding error probability at the cost of higher time complexity. Given the value $P\in[n]$, we pick $P$ permutations from the set $\{\sigma_0,\sigma_1,\dots,\sigma_{n-1}\}$. Simulation results indicate that the performance of our decoder does not depend on which $P$ permutations we choose, so we can simply pick the permutations $\sigma_0,\sigma_1,\dots,\sigma_{P-1}$. Then we build a large parity check matrix $H$ of size $P(n-1)\times n$, which consists of all the row vectors of $H_0,H_1,\dots,H_{P-1}$. Our neural decoder performs Belief Propagation on the Tanner graph of $H$.

Let us now take a closer look at the matrix in the bottom-left corner of Fig.~\ref{fig:illus}. Since the last $7$ columns of this matrix are cyclic shifts of each other, it is natural to impose a shift invariant structure on weights of the neural decoder associated with the Tanner graph of this matrix. In general, given a RM code or an extended BCH code $\cC$, the last $(n-1)$ columns of $H_0$ are also cyclic shifts of each other, and we also impose the shift-invariant structure on the weights of the neural decoder. Note that such a structure was already adopted in \cite{Chen21}. The major innovation of our new decoder is that we further extend this invariant structure to the columns obtained by the permutations $\sigma_0,\sigma_1,\dots,\sigma_{n-1}$.

We are now ready to formally define our decoder. We index the columns of $H_0$ from $0$ to $n-1$. Suppose that the number of $1$'s in the first column of $H_0$ is $u$, and let $\{i_1,i_2,\dots,i_u\}\subseteq [n]$ be the set satisfying that the $(i_b,1)$th entry of $H_0$ is $1$ for all $b\in[u]$. Let $\pi_j$ be the permutation obtained by $j-1$ right cyclic shifts on the set $\{1,2,\dots,n-1\}$. 
We use a triple $(z,i,j)$ to denote an edge in the Tanner graph of $H$. Recall that $H$ contains all the row vectors of $H_0,H_1,\dots,H_{P-1}$. The edge $(z,i,j)$ corresponds to the $(i,j)$th entry of the matrix $H_z$.
Then $(0,\pi_j(i_1), j), (0,\pi_j(i_2), j), \dots, (0,\pi_j(i_u), j)$ are the $u$ edges that contain $v_j$ as an endpoint in the $j$th column of $H_0$; see Fig.~\ref{fig:illus}. For an edge $e$ in the Tanner graph, we use $x^{[s]}(e)$ to denote the message on $e$ in the $s$-th iteration. In the calculations of each odd iteration, we use the following $u^2$ weights: $\{w_{b,b'}^{[s]}:b,b'\in[u],b\neq b'\}$ and $\{w_b^{[s]}:b\in[u]\}$. 
For odd $s$ and an edge $e=(z,\pi_{\sigma_z(j)}(i_b), j)$ with $b\in[u]$, the message $x^{[s]}(e)$ is given by
\begin{align}
 x^{[s]}(e)=
x^{[s]} & ((z,\pi_{\sigma_z(j)}(i_b), j))  
= \tanh \Big( \frac{1}{2} \Big( w_b^{[s]} L_j
\label{eq:ourodd} \\
& + \sum_{b'\in [u]\setminus \{b\} }
w_{b', b}^{[s]} ~
x^{[s-1]}((z,\pi_{\sigma_z(j)}(i_{b'}), j)) \Big) \Big) .
\nonumber
\end{align}
The calculations of even iterations are the same as those in the vanilla BP algorithm: For even $s$ and an edge $e=(z, i, j)$,
$$
x^{[s]}(e) 
= 2 \tanh^{-1} \Big( \prod_{e'\in N_z(c_i)\setminus \{e\} } x^{[s-1]}(e') \Big) ,
$$
where $N_z(c_i)$ is the set of all the edges containing $c_i$ as an endpoint in $H_z$.
In the calculations of the output layer, we use the following $u$ weights: $\{w_b^{\out}:b\in[u]\}$. The $j$th output is given by
\begin{equation} \label{eq:ourout}
o_j = L_j  + \sum_{z=0}^{P-1} \sum_{b=1}^u 
w_b^{\out} ~
x^{[2t]}((z,\pi_{\sigma_z(j)}(i_b), j)) 
\end{equation}
for $j\in[n]$, where $2t$ is the total number of iterations.

\section{Simulation results}

We present the simulation results of our new decoder in this section, and we compare its performance with the decoders in \cite{Nachmani18,Chen21}. In particular, we refer to the neural decoder in \cite{Nachmani18} as N\_18. We refer to the cyclically equivariant neural decoder in \cite{Chen21} as Cyc, and the list decoding version of Cyc is referred to as Cyc\_list with a parameter $\ell$ specifying the list size. Our decoder also has a parameter $P$ which is the number of permutations we use in the decoder. Note that when $P=1$, our new decoder reduces to the Cyc decoder in \cite{Chen21}.

\begin{figure}
\centering
\begin{subfigure}{0.9\linewidth}
\centering
\includegraphics[width=\textwidth]{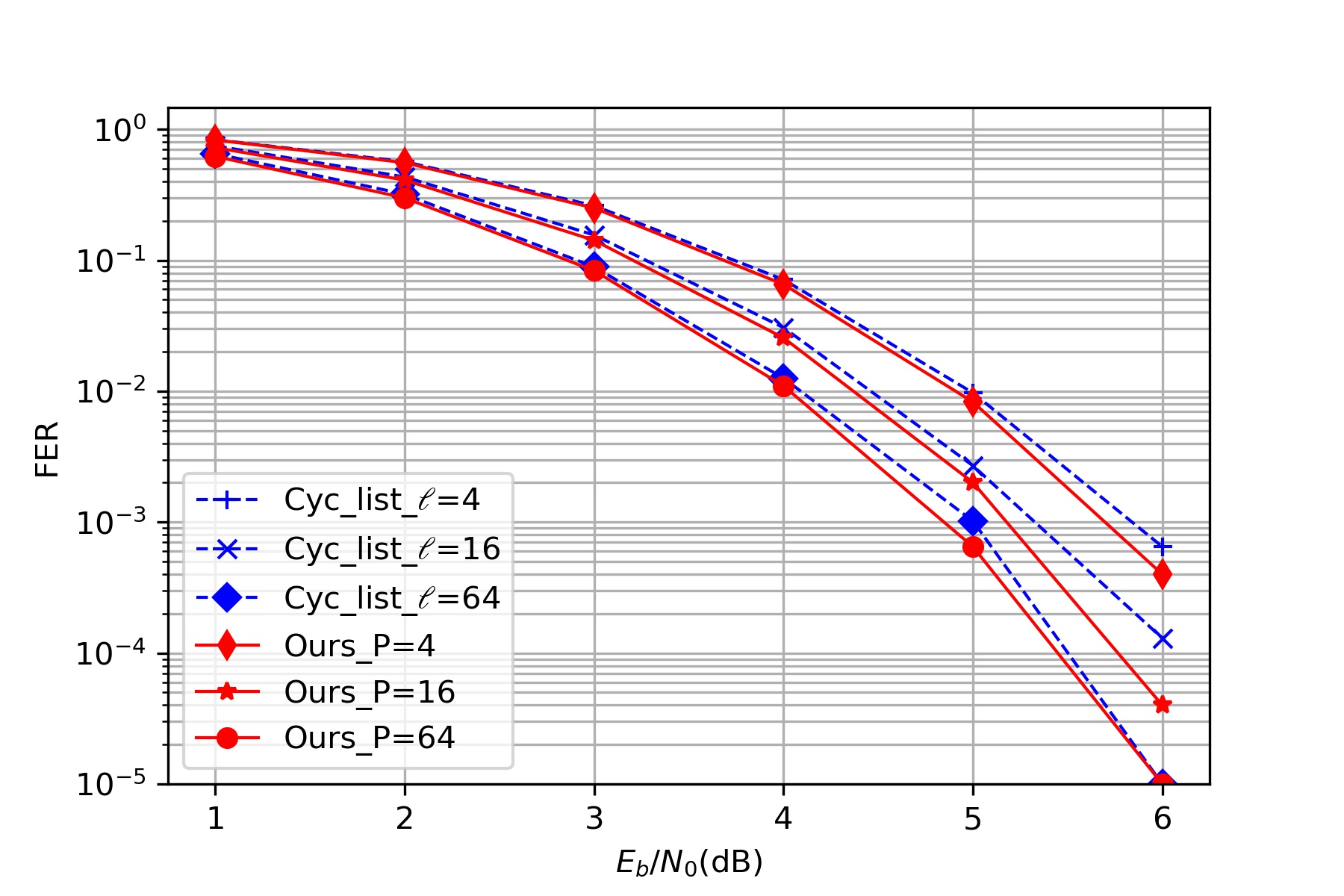} 
\caption{BCH(63,36)}
\end{subfigure}
\\
\begin{subfigure}{0.9\linewidth}
\centering
\includegraphics[width=\textwidth]{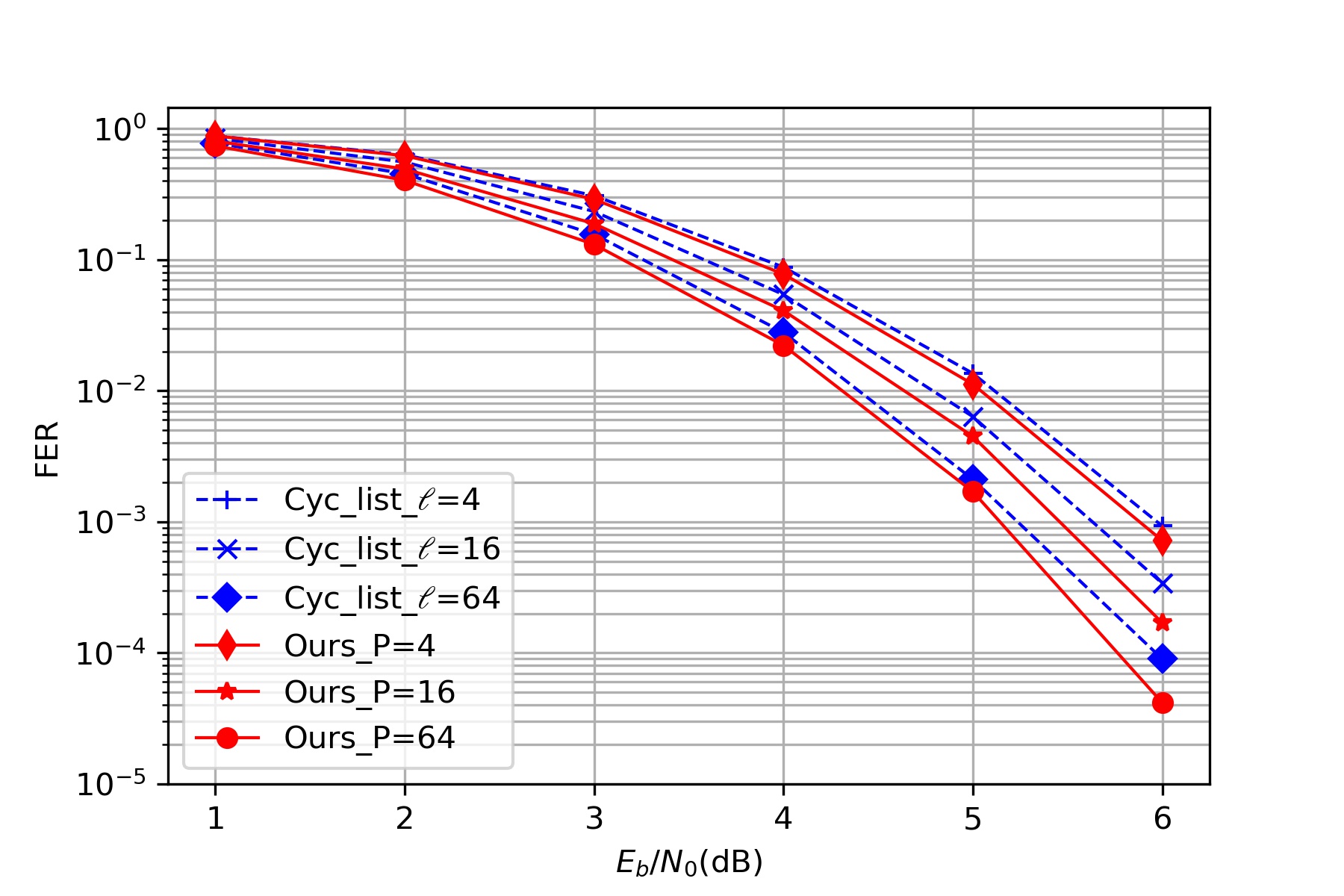} 
\caption{BCH(63,45)}
\end{subfigure}
\caption{Comparison of FER between our new decoder and Cyc\_list. When $P=\ell$, our decoder always has the same or even smaller FER than Cyc\_list.}
\label{fig:FER}
\end{figure}

\begin{figure*}
\centering
\begin{subfigure}{0.47\textwidth}
\centering
\includegraphics[width=\textwidth]{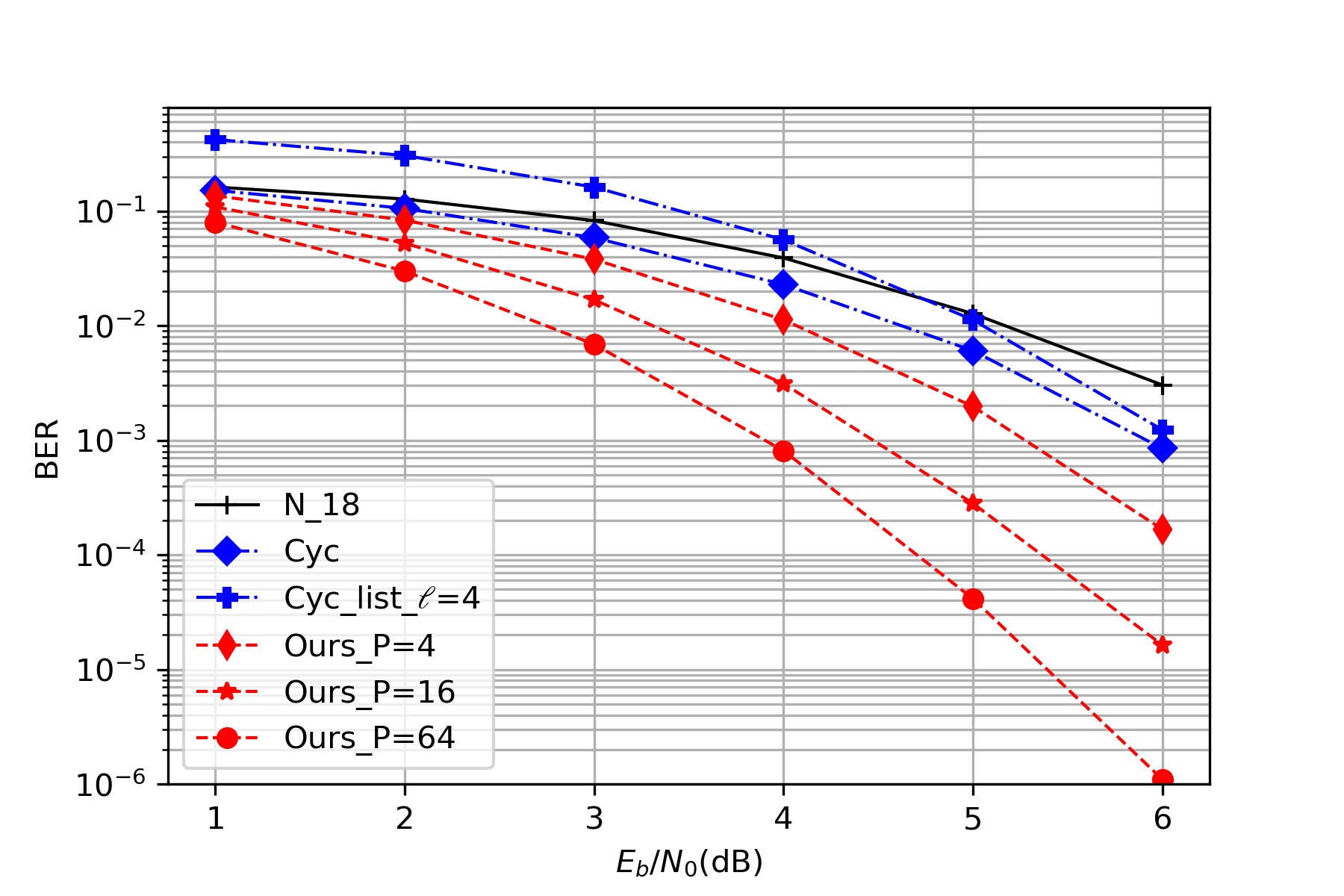} 
\caption{BCH(63,24)}
\end{subfigure}
~
\begin{subfigure}{0.47\textwidth}
\centering
\includegraphics[width=\textwidth]{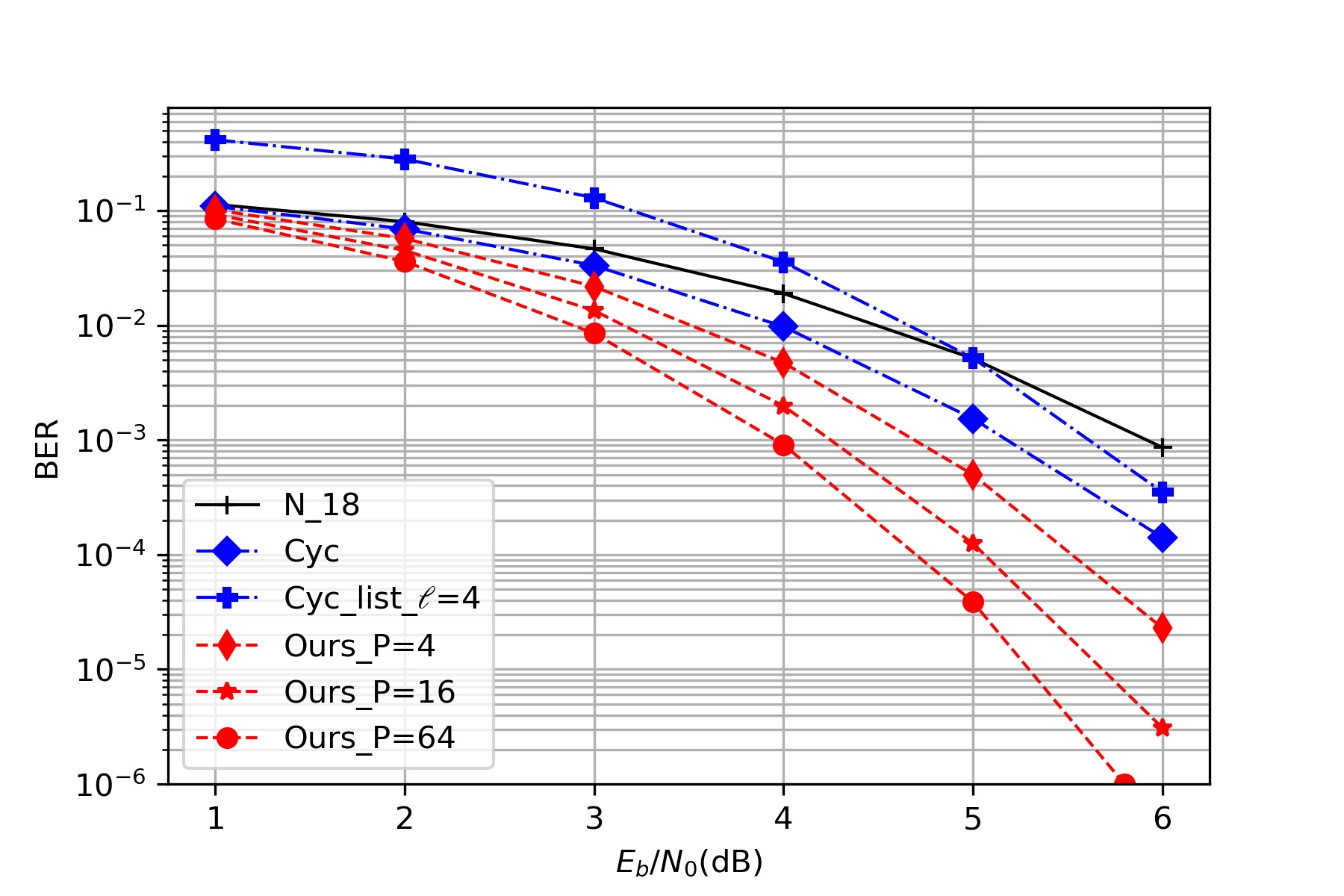} 
\caption{BCH(63,36)}
\end{subfigure}

\begin{subfigure}{0.47\textwidth}
\centering
\includegraphics[width=\textwidth]{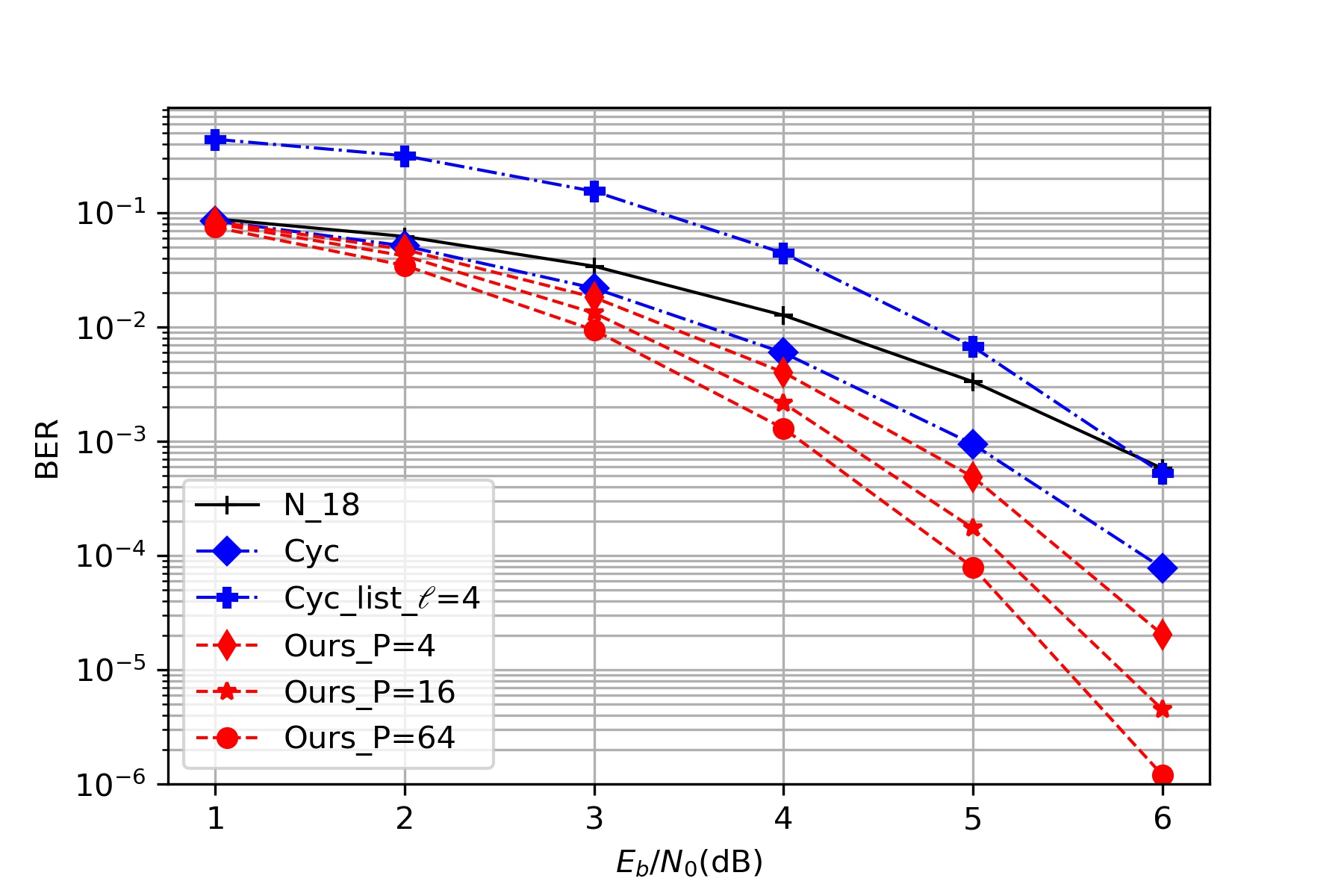} 
\caption{BCH(63,45)}
\end{subfigure}
~
\begin{subfigure}{0.47\textwidth}
\centering
\includegraphics[width=\textwidth]{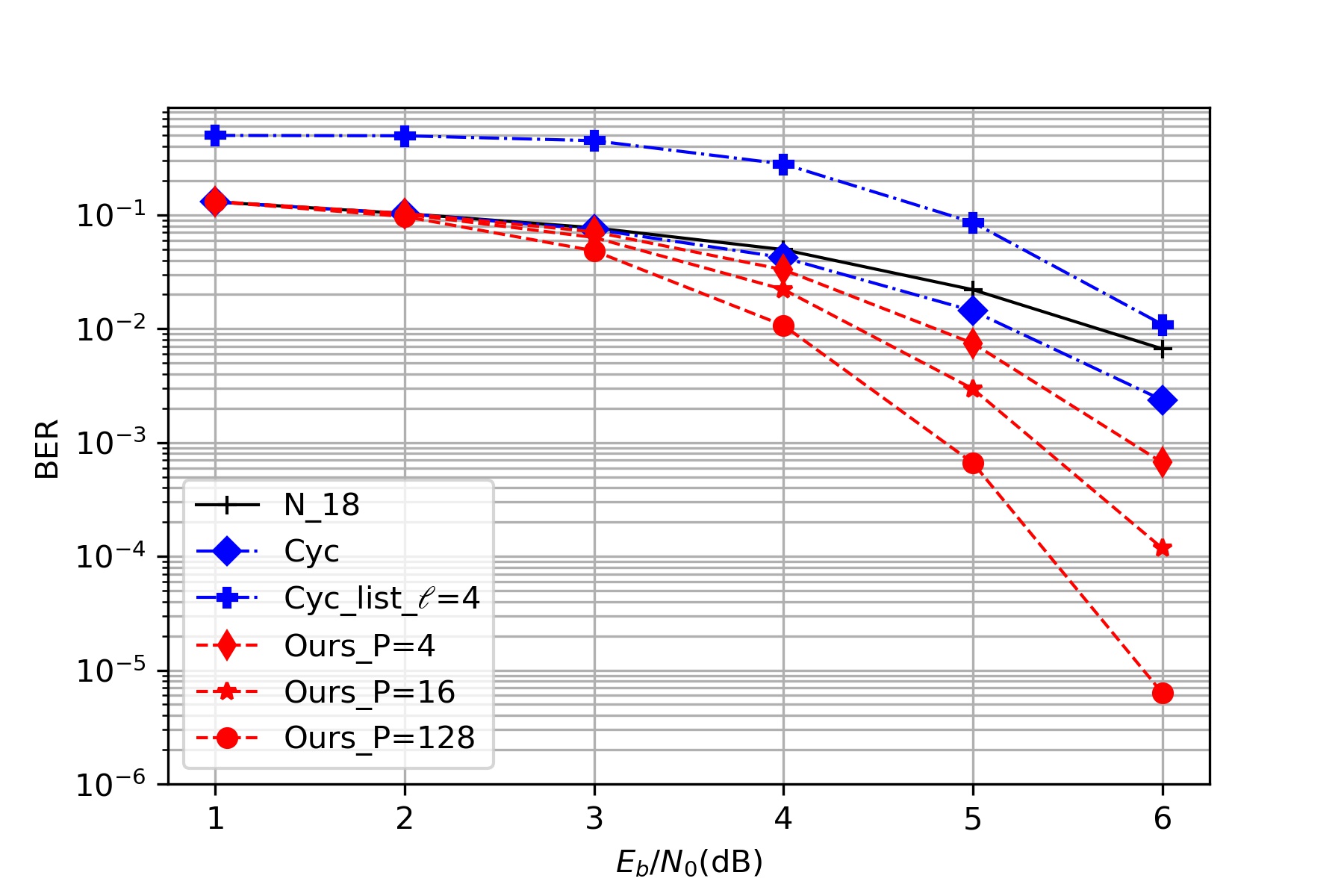} 
\caption{BCH(127,64)}
\end{subfigure}

\begin{subfigure}{0.47\textwidth}
\centering
\includegraphics[width=\textwidth]{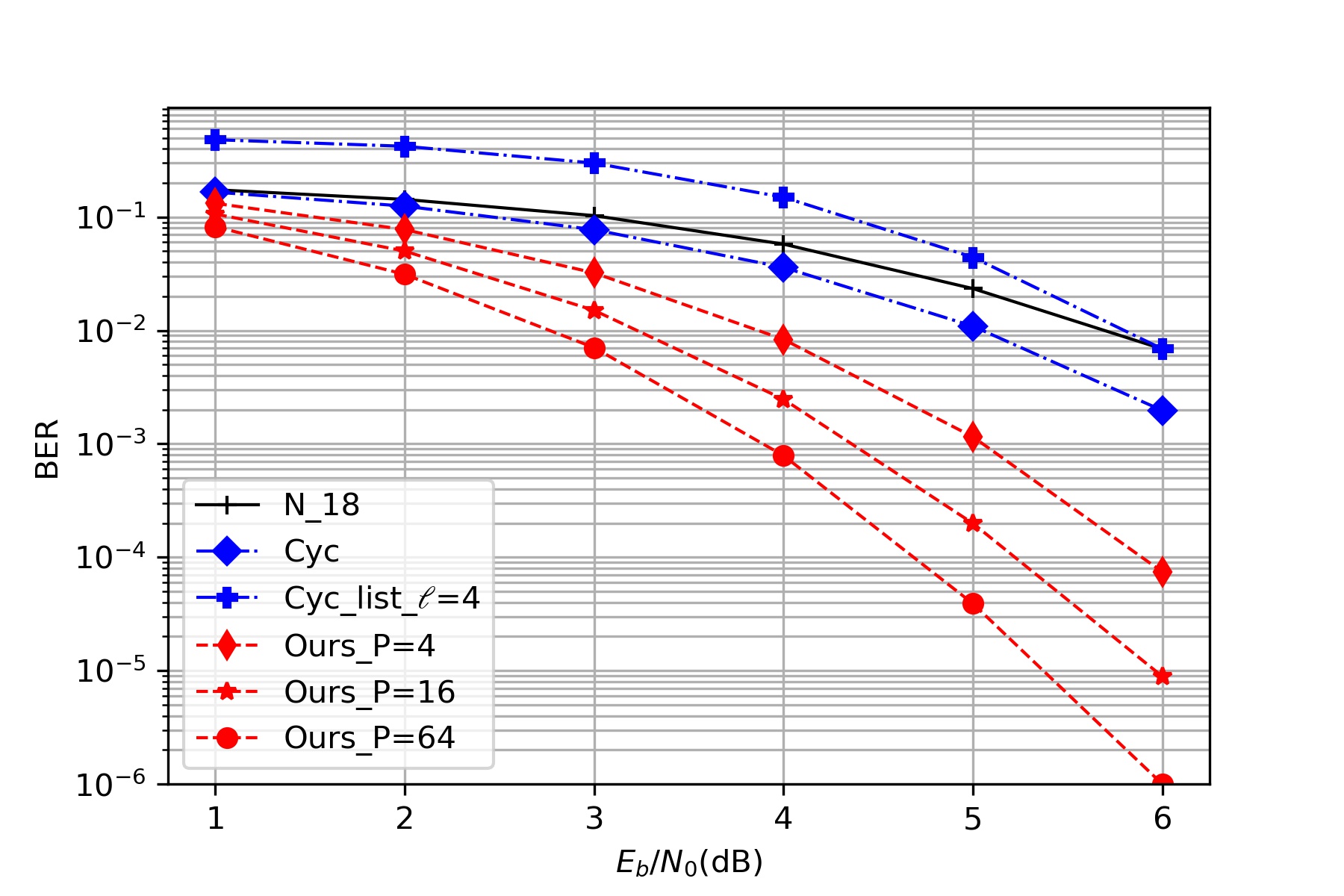} 
\caption{Punctured RM(63,22)}
\end{subfigure}
~
\begin{subfigure}{0.47\textwidth}
\centering
\includegraphics[width=\textwidth]{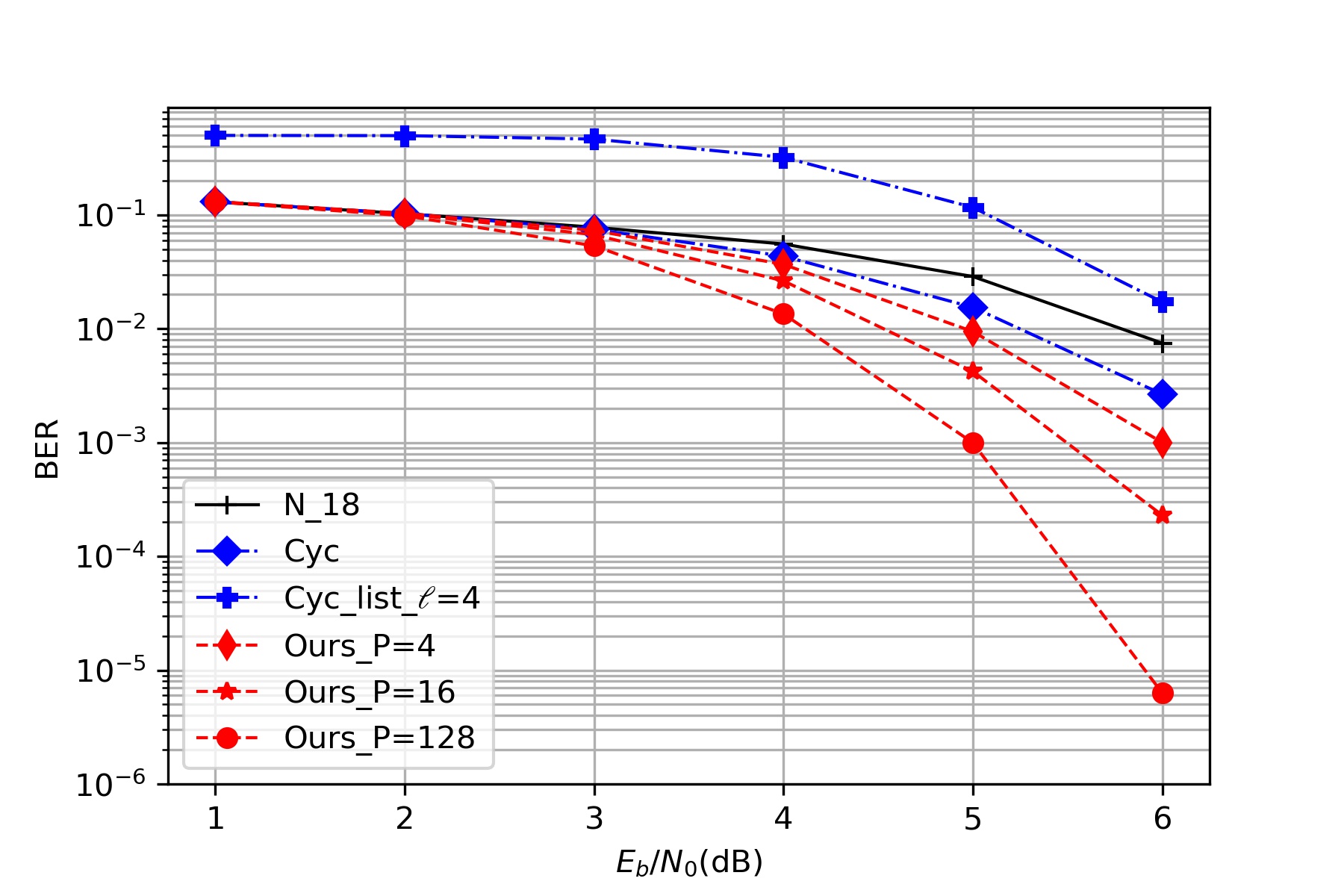} 
\caption{Punctured RM(127,64)}
\end{subfigure}

\caption{N\_18 refers to the neural decoder in \cite{Nachmani18}. Cyc and Cyc\_list refer to the decoders in \cite{Chen21}, where the parameter $\ell$ is the list size. For our new decoder, the parameter $P$ is the number of permutations used in the decoder. When $P=1$, our new decoder reduces to the Cyc decoder in \cite{Chen21}.
As we increase the value of $P$, our decoder gracefully reduces the BER. When we take $\ell=P=4$, our decoder has 1 to 2dB gain over the Cyc\_list decoder in terms of BER.}
\label{fig:simu}
\end{figure*}

When we set $P=\ell$ in our decoder and Cyc\_list, the BER of our decoder demonstrates 1 to 2 dB improvements over Cyc\_list (see Fig.~\ref{fig:simu}), and the FER of our decoder remains the same or even smaller (see Fig.~\ref{fig:FER}). Moreover, our decoder also reduces the running time by at least $15\%$; see Table~\ref{tb:time}. An important advantage of our decoder is that when we increase the value of $P$, our decoder gracefully reduces the BER. In contrast, when $\ell$ takes a small value, e.g., $\ell=4$, the BER of Cyc\_list is even worse than Cyc while the running time is 4 times larger; see Fig.~\ref{fig:simu}. As explained in the Introduction, this is because the Cyc\_list decoder does not optimize the fraction of incorrect bits in the incorrectly decoded codewords. Note that this fraction is precisely the ratio between BER and FER. Whenever Cyc\_list outputs an incorrect codeword, it simply picks a random one, so with high probability the fraction of incorrect bits in this randomly chosen codeword is very close to $1/2$, as indicated by Table~\ref{tb:ratio}. In contrast, our decoder always tries to minimize the fraction of incorrect bits even when it outputs the wrong decoding result, and the fraction of incorrect bits in the wrong decoding result is smaller than $0.1$ for our decoder; see Table~\ref{tb:ratio}.

\begin{table}
\centering
    \caption{The BER/FER for three SNR values and different decoders.
    This ratio is equal to the fraction of incorrect bits in the incorrectly decoded codewords. The list size in Cyc\_list \cite{Chen21} is $\ell=4$, and the number of permutations used in our decoder is $P=4$.}
    \label{tb:ratio}
    \begin{tabular}{lcccccc}
    \toprule
    Code & \multicolumn{3}{c}{BCH(63,36)} & \multicolumn{3}{c}{BCH(63,45)} \\
    \cmidrule(r){2-4}   \cmidrule(r){5-7} 
    Decoder/SNR & 4  & 5 & 6 & 4 & 5 & 6 \\
    \midrule
    Cyc\_list & 0.500 & 0.499 & 0.503 & 0.499 & 0.499 & 0.496 \\
    Ours & 0.072 & 0.060 & 0.058 & 0.056 & 0.047 & 0.033 \\ \bottomrule
\end{tabular}
\end{table}

\begin{table*}
\centering
    \caption{Comparison between the decoding time of Cyc\_list in \cite{Chen21} and our new decoder. When $P=\ell$, our new decoder reduces the running time by at least $15\%$.}
    \label{tb:time}
    \resizebox{\textwidth}{!}{
    \begin{tabular}{cccccccc}
    \toprule
    Code       & Cyc    & Cyc\_list\_$\ell=4$ & Ours\_$P=4$ & Cyc\_list\_$\ell=16$ & Ours\_$P=16$ & Cyc\_list\_$\ell=64$ & Ours\_$P=64$ \\
    \midrule
    BCH(63,36) & 3.59ms & 17.4ms  & 15.2ms & 68.6ms  & 56.7ms  & 268ms & 213ms  \\
    BCH(63,45) & 4.60ms & 22.4ms  & 18.9ms   & 85.2ms  & 72.5ms & 343ms & 289ms\\
    \bottomrule
    \end{tabular}}
\end{table*}

\bibliographystyle{IEEEtran}
\bibliography{improve}

\end{document}